\documentclass[english,prl,twocolumn,superscriptaddress,letterpaper]{revtex4}

\usepackage{ae} 
\usepackage[T1]{fontenc}
\usepackage[ansinew]{inputenc}
\usepackage{amsmath}
\usepackage{amssymb}
\usepackage{dsfont}
\usepackage{graphicx}
\usepackage{color}
\usepackage{epsfig}
\usepackage{booktabs}

\def\eps{\epsilon}
\def\d{{\rm d}}
\newcommand{\Cut}[0]{\operatorname{Cut}}
\newcommand\PV{\operatorname{PV}}
\renewcommand\Im{\operatorname{\mathrm{Im}}}

\usepackage{varioref}
\usepackage{makeidx}
\makeindex

\usepackage[english]{babel}

\begin{document}

\title{Imaginary Parts and Discontinuities of Wilson Line Correlators}

\author{Eric Laenen}
\affiliation{Nikhef, Science Park 105, 1098 XG Amsterdam, The Netherlands}
\affiliation{ITFA, University of Amsterdam, Science Park 904, 1018 XE Amsterdam, The Netherlands}
\affiliation{ITF, Utrecht University, Leuvenlaan 4,  3584 CE Utrecht, The Netherlands}
\author{Kasper J. Larsen}
\affiliation{Nikhef, Science Park 105, 1098 XG Amsterdam, The Netherlands}
\affiliation{Institute for Theoretical Physics, ETH Z{\"u}rich, 8093 Z{\"u}rich, Switzerland}
\affiliation{School of Natural Sciences, Institute for Advanced Study, Princeton, NJ 08540, USA}
\author{Robbert Rietkerk}
\affiliation{Nikhef, Science Park 105, 1098 XG Amsterdam, The Netherlands}
\affiliation{ITFA, University of Amsterdam, Science Park 904, 1018 XE Amsterdam, The Netherlands}

\begin{abstract}
We introduce a notion of position-space cuts of eikonal diagrams,
the set of diagrams appearing in the perturbative expansion of the correlator of a set of straight
semi-infinite Wilson lines. The cuts are applied
directly to the position-space representation of any such diagram and
compute its imaginary part to the leading order in the dimensional
regulator. Our cutting prescription thus defines a position-space
analog of the standard momentum-space Cutkosky rules.
Unlike momentum-space cuts which put internal lines
on shell, position-space cuts constrain a number of the gauge bosons
exchanged between the energetic partons to be lightlike,
leading to a vanishing and a non-vanishing imaginary part
for space- and timelike kinematics, respectively.
\end{abstract}

\maketitle

\section{Introduction}

The infrared singularities of gauge theory scattering amplitudes
play a fundamental role in particle physics for both phenomenological
and theoretical reasons. Knowing the structure of long-distance
singularities is necessary for combining the real and virtual contributions
to the cross section, as the divergences of the separate contributions
only cancel once they are added. In addition, infrared singularities
dictate the structure of large logarithmic contributions
to the cross section, allowing such terms to be resummed.
Long-distance singularities, moreover, have several highly interesting properties.
They have a universal structure among different gauge theories;
their exponentiation properties \cite{Yennie:1961ad,Gatheral:1983cz,Frenkel:1984pz,Gardi:2010rn,Mitov:2010rp,Gardi:2013ita}
and their relation to the renormalization of Wilson line correlators
\cite{Korchemsky:1985xj,Korchemsky:1987wg}
allow the exploration of the all-order structure of their perturbative expansion,
a feat currently unattainable for complete scattering amplitudes.

The key tool for computing the infrared singularities of
scattering amplitudes is provided by the eikonal approximation
in which each parton $i$ emerging from the hard scattering
acts as a source of soft gluon radiation and
is replaced by a semi-infinite path ordered Wilson line
\begin{equation}
\Phi_{v_i} \equiv \mathcal{P}
\exp\left( ig \int_0^\infty \hspace{-0.6mm} \d t \hspace{0.9mm}
v_i \cdot A(tv_i) \right) \,,
\label{eq:def_of_Wilson_line}
\end{equation}
which extends from time $t=0$, when the hard scattering takes place,
to infinity along the classical trajectory of the hard parton,
spanned by its four-velocity $v_i^\mu$. The long-distance singularities
of the scattering amplitude of the hard partons are then given by
the eikonal amplitude
\begin{equation}
\mathcal{S}(v_i \cdot v_j, \eps) \equiv
\langle 0 \hspace{0.1mm} | \hspace{0.1mm} \Phi_{v_1} \otimes \Phi_{v_2}
\otimes \cdots \otimes \Phi_{v_n} \hspace{0.1mm} | \hspace{0.1mm} 0 \rangle \,,
\label{eq:Wilson_line_correlator_def}
\end{equation}
which has the exact same soft singularities as the original full
amplitude, but is much simpler to compute. Owing to the scale
invariance of the Wilson line correlator (\ref{eq:Wilson_line_correlator_def}),
its infrared singularities can be computed equivalently
by studying its ultraviolet renormalization factor
\cite{Korchemsky:1985xj,Korchemsky:1987wg}.
This renormalization factor forms a matrix in the space of available
color configurations, called the soft anomalous dimension matrix. This matrix has been
computed through two loops for massless \cite{Aybat:2006wq,Aybat:2006mz}
as well as massive \cite{Becher:2009kw,Ferroglia:2009ep,Ferroglia:2009ii,Mitov:2010xw}
Wilson lines, and there has been recent progress toward the
three-loop result in Refs.~\cite{Henn:2013wfa,Gardi:2013saa,Falcioni:2014pka}.
In processes involving only two Wilson lines, the soft matrix
reduces to the cusp anomalous dimension, which has been computed
in QCD up to three loops \cite{Korchemsky:1987wg,Kidonakis:2009ev,Grozin:2014hna}.

In this Letter we introduce a notion of cuts of eikonal diagrams---i.e.,
the diagrams contributing to the eikonal amplitude. Applied to
any eikonal diagram, the cuts produce its discontinuities,
in analogy with the Cutkosky rules for standard Feynman
diagrams. The discontinui\-ties are in turn readily combined
to yield the imaginary part of the diagram. A direct computational
method of the latter is desirable in a variety of contexts,
e.g. rapidity gaps \cite{Forshaw:2006fk,Forshaw:2008cq},
cross section calculations \cite{Korchemsky:1992xv,Gardi:2005yi},
and the breaking of collinear factorization theorems caused by exchanges
of Glauber-region (i.e., transverse) gluons \cite{Catani:2011st,Forshaw:2012bi}.
Regarding the latter, the resulting factorization-breaking
terms are purely imaginary and take the form of the non-Abelian analog
of the QED Coulomb phase. By utilizing the all-order exponentiation
property of the eikonal amplitude, the latter could be obtained
by computing the imaginary part of the exponent.
Introducing cuts of eikonal diagrams can also be viewed as
the first step toward extending the modern unitarity
method~\cite{Bern:1994zx,Bern:1994cg,Bern:1996je,Britto:2004nc,Britto:2005ha,Forde:2007mi,Mastrolia:2009dr,Kosower:2011ty,CaronHuot:2012ab}
to Wilson line correlators. In unitarity,
loop-level (non-eikonal) amplitudes are computed by expanding
the amplitude in a basis of integrals and determining the basis
coefficients by taking cuts which measure the discontinuity of
the amplitude in its various kinematic channels.

We emphasize that a cutting prescription acting on the
momentum-space representation of eikonal diagrams was defined in Ref.~\cite{Korchemsky:1987wg}.
In contrast, the cuts introduced here
are applied to the position-space representation of the diagrams.
As we shall see, position-space cuts offer a substantial simplification
over momentum-space cuts in the computation of imaginary parts
of eikonal diagrams.

\section{Imaginary parts and their physical origin}

In this section we discuss the origin of the imaginary part of
Wilson line correlators from the point of view of causality
as well as unitarity.

We will adopt the convention that all velocities are outgoing.
Ultraviolet divergences are regulated by computing
all diagrams in $D=4-2\eps$ dimensions with $\eps > 0$.
To avoid complications arising from regulating collinear singularities,
we take all velocities to be timelike, $v_i^2 = 1$.
For notational convenience, we will drop color factors,
coupling constants and factors of $\frac{\Gamma(D/2 -1)}{4\pi^{D/2}}$.

We will start our investigations by examining
the simplest eikonal diagram, the one-loop exchange.
The position-space representation of this diagram
is obtained by direct perturbative expansion of
Eq.~(\ref{eq:Wilson_line_correlator_def}), yielding
\begin{equation}
F^{(1)} = \mu^{2\epsilon} \int_0^\infty \hspace{-0.8mm}
\int_0^\infty \frac{\d t_1 \hspace{0.3mm} \d t_2 \hspace{0.7mm} v_1 \cdot v_2}
{\big[ {-}(t_1 v_1 - t_2 v_2)^2 + i\eta \big]^{1-\epsilon}} \,,
\label{eq:one-loop_diagram_x-space}
\end{equation}
where $t_1, t_2$ have the dimension of time and denote the positions of
the attachment points of the soft-gluon propagator on the
Wilson lines spanned by the four-velocities $v_1$ and $v_2$.
The integrations in Eq.~(\ref{eq:one-loop_diagram_x-space}) yield
an infrared divergence which can be extracted via the change of variables
$(t_1, t_2)= (\lambda x, \lambda(1-x))$ with $0\leq x \leq 1$,
where $\lambda$ has the dimension of length,
\begin{equation}
F^{(1)}\hspace{-0.8mm}= \mu^{2\epsilon} \hspace{-0.9mm} \int_0^\infty
\hspace{-1.3mm} \frac{\d \lambda \hspace{0.7mm} e^{-\Lambda \lambda}}{\lambda^{1-2\epsilon}}
\hspace{-0.7mm} \int_0^1 \frac{\d x \hspace{0.8mm} v_1 \cdot v_2}
{\big[{-}(x v_1 \hspace{-0.3mm}-\hspace{-0.3mm} (1\hspace{-0.3mm}-\hspace{-0.3mm}x) v_2)^2
\hspace{-0.3mm}+\hspace{-0.3mm} i\eta \big]^{1-\epsilon}} \,,
\label{eq:one-loop_diagram_x-space_lambda_s-channel}
\end{equation}
where the infrared divergence arising from the exchange of
gluons of increasingly longer wavelength is regularized
by the exponential damping factor $e^{-\Lambda \lambda}$
with $\Lambda \ll 1$.

The diagram is then readily evaluated, yielding
in, respectively, time- and spacelike kinematics to $\mathcal{O}(\eps^{-1})$,
\begin{equation}
F^{(1)} \hspace{-0.5mm}=\hspace{-0.5mm}
\frac{1}{2\eps} \left( \frac{\mu}{\Lambda} \right)^{2\eps}
\hspace{-0.5mm} \times \hspace{0.5mm} \left\{ \hspace{-0.7mm}
\begin{array}{cllr}
(\gamma - \pi i) \coth \gamma \hspace{2mm} &\mathrm{for} & v_1 \cdot v_2 &\hspace{-1mm}> 0 \,\phantom{.} \\[2.5mm]
\gamma \coth \gamma \hspace{2mm} &\mathrm{for} & v_1 \cdot v_2 &\hspace{-1mm}< 0 \,,
\end{array} \right.
\label{eq:one-loop_diagram_result}
\end{equation}
where the angle $\gamma$ is defined through
$\cosh \gamma \equiv |v_1 \cdot v_2|$, and where, e.g.,
the timelike result can be obtained from the spacelike
one by the analytic continuation $\gamma \to \pi i - \gamma$.

We observe that the imaginary part of the one-loop diagram
in Eq.~(\ref{eq:one-loop_diagram_result}) is, respectively,
non-vanishing and vanishing. From the position-space representation
(\ref{eq:one-loop_diagram_x-space}) of the diagram,
the origin of the imaginary part can be understood from a
simple causality consideration as follows. (As our focus is on
computing the imaginary part to the leading order in $\eps$,
the $\eps$ in the propagator exponent can be dropped once
the infrared divergence has been extracted.)
For timelike kinematics $v_1 \cdot v_2 >0$,
there are regions $\frac{t_1}{t_2} = e^{\pm \gamma}$ within the
integration domain where $(t_1 v_1 - t_2 v_2)^2 =0$,
so that the $-i\eta$ term becomes relevant and generates
an imaginary part. Physically, what is happening at
such times $t_1, t_2$ is that the two partons traveling
along $v_1$ and $v_2$ become lightlike separated. As a result, the
phases of their states can be changed through
the exchange of lightlike gluons (or photons).
In contrast, for spacelike kinematics $v_1 \cdot v_2 <0$,
the integral in Eq.~(\ref{eq:one-loop_diagram_x-space})
has a vanishing imaginary part: the denominator
$(t_1 v_1 - t_2 v_2)^2$ is strictly positive within the region
of integration, and the $-i\eta$ can thus be dropped.
In this case, the partons are never lightlike separated,
and the phases of their states cannot be changed
through the exchange of lightlike massless gauge bosons.

These observations on the evolution of the phases of the hard-parton
states, related in the interaction picture through time
evolution by $|f\rangle_I = e^{i \int_0^{1/\Lambda}
\d t \hspace{0.6mm} V_I t} |i\rangle_I$, suggest that
the imaginary part of the anomalous dimension of the correlator of
two Wilson lines defines an interparton potential.
This is indeed the case in conformal gauge theories
owing to the state-operator correspondence \cite{Chien:2011wz};
in QCD, the relation holds up to terms proportional
to the beta function \cite{Grozin:2014hna}.

Let us turn to the complementary question of
how the imaginary part of the one-loop diagram
may be obtained from its momentum-space representation,
\begin{equation}
F^{(1)} \hspace{-0.6mm}=\hspace{-0.2mm} i \mu^{2\epsilon}
\hspace{-1.2mm} \int \hspace{-0.9mm} \frac{\d^D k}{(2\pi)^D}
\frac{v_1 \cdot v_2}{(k^2 \hspace{-0.3mm}+\hspace{-0.3mm} i\eta)
(v_1 \hspace{-0.2mm} \cdot \hspace{-0.2mm} k \hspace{-0.3mm}+\hspace{-0.3mm} i\eta)
(v_2 \hspace{-0.2mm} \cdot \hspace{-0.2mm} k \hspace{-0.3mm}-\hspace{-0.3mm} i\eta)} \,.
\label{eq:one-loop_diagram_p-space}
\end{equation}
Such a cutting prescription was provided in Ref.~\cite{Korchemsky:1987wg}
where it was shown that the imaginary part of the one-loop diagram
in Eq.~(\ref{eq:one-loop_diagram_p-space}) is obtained by
replacing the two eikonal propagators by delta functions,
\begin{align}
2i \hspace{0.4mm} \mathrm{Im} \hspace{0.6mm} F^{(1)} &= (2\pi)^2 i \hspace{0.6mm}
\theta (v_1^0) \theta (v_2^0) (v_1 \cdot v_2) \mu^{2\epsilon} \nonumber\\
&\hspace{7mm}\times \int \hspace{-0.5mm} \frac{\d^D k}{(2\pi)^D}
\frac{\delta(v_1 \cdot k) \delta(v_2 \cdot k)}{k^2 + i\eta} \,.
\label{eq:map_p-space_cut_to_x-space_1}
\end{align}
From this representation we observe that the support of the delta
functions in Eq.~(\ref{eq:map_p-space_cut_to_x-space_1})
is the region where the momentum of the exchanged gluon
is maximally transverse, $v_i \cdot k \approx 0$,
which was identified in Ref.~\cite{Korchemsky:1987wg} as
the Glauber region \cite{Collins:1983ju}. Furthermore,
the delta functions have the effect of putting the
hard partons on shell, making them asymptotic states.
Thus, in momentum space, the imaginary part
arises from the two hard partons going on shell and exchanging
Glauber gluons.

We conclude that the position- and momentum-space representations
of eikonal diagrams offer complementary points of view
on the origin of their imaginary part, based, respectively,
on causality and unitarity considerations.

The momentum-space cuts applied in
Eq.~(\ref{eq:map_p-space_cut_to_x-space_1}) have the conceptual advantage
of factoring eikonal diagrams into on-shell lower-loop and tree diagrams
which can be computed as independent objects. However, at $L$ loops,
the resulting cut diagrams involve integrals over up to $(L+1)$-particle phase space.
The evaluation of these integrals poses a substantial
computational challenge which limits the applicability
of the momentum-space cutting prescription
for obtaining imaginary parts.

\section{Cuts of eikonal diagrams without internal vertices}

In this section we will derive a formula for the imaginary part
of $L$-loop eikonal diagrams containing no internal (i.e., three-
or four-gluon) vertices to
the leading order in the dimensional regulator $\epsilon$.
We will interchangeably refer to such diagrams as ladder-type diagrams.
The basic observation is that in position space
these diagrams are iterated integrals, and thus their imaginary part
can be obtained by decomposing the real-line integrations into
principal-value and delta function contributions.

In position space, an arbitrary $L$-loop ladder-type eikonal diagram
consists of $L$ soft-gluon propagators, or \emph{rungs}.
Each rung extends between the eikonal lines
spanned by any two (possibly identical) external four-velocities
$v_1, \ldots, v_n$ where $1 \leq n \leq L+1$.
For the $j$th rung we will denote these four-velocities
by $v_{\ell_j}$ and $v_{r_j}$. Let $t_{i,k}$ denote the position
of the $k$th attachment on the eikonal line spanned by $v_i$, counting
from the hard interaction vertex and outwards so that
$0\leq t_{i,1} < \cdots < t_{i,N_i}$ where
$N_i$ denotes the total number of soft-gluon attachments on the eikonal
line. Furthermore, for the $j$th rung, we will let the variables
$m_j$ and $n_j$ record the soft-gluon attachment numbers on
the eikonal lines spanned by $v_{\ell_j}$ and $v_{r_j}$, respectively.
The $L$-loop eikonal diagram is then defined as the $2L$-fold integral
\begin{align}
F^{(L)} &= \mu^{2L\epsilon} \prod_{j=1}^L \int_0^\infty
\frac{\d t_{\ell_j, m_j} \d t_{r_j, n_j} (v_{\ell_j} \cdot v_{r_j})}
{\big[ {-}(t_{\ell_j, m_j} v_{\ell_j} - t_{r_j, n_j} v_{r_j})^2
+ i\eta \big]^{1-\epsilon}} \nonumber \\
&\hspace{18mm}\times \prod_{i=1}^n \prod_{k=0}^{N_i}
\theta (t_{i,k+1} - t_{i,k}) \,,
\label{eq:L-loop_ladder_diagram_def}
\end{align}
where it is implied that $t_{i,N_i +1} \equiv \infty$ and $t_{i,0} \equiv 0$.
Without loss of generality, we assume that no rungs attach with
both end points to the same Wilson line. In such diagrams these
rungs can be integrated out, each producing to leading order in
$\epsilon$ a factor of $1/\epsilon$ times the diagram without
these rungs~\footnote{More explicitly, these integrations will
produce a factor of $1/\epsilon$ times epsilonic powers of the
attachment points of rungs attached in between the end points of
the rung being integrated out.}. For the latter we can then use our formalism.

\begin{figure*}[!t]
\begin{center}
\includegraphics[width=0.90\textwidth]{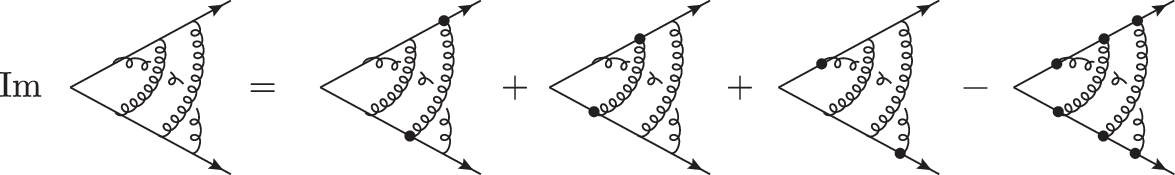}
{\vskip 1mm}
\caption{Schematic illustration of the formula~(\ref{eq:Master Im F})
for the imaginary part of an eikonal diagram with no internal vertices.
The black dots at the end points of a soft-gluon propagator indicate
that the propagator has been cut; i.e., replaced by a delta function.
Physically, the black dots represent the emission and absorption
of a lightlike gauge boson. It is implied that the integrals over
the attachment points of uncut soft propagators are principal-value integrals.}
\label{fig:Im_of_three-loop_ladder}
\end{center}
\end{figure*}

In order to extract the imaginary part of $F^{(L)}$ from the integral
representation in Eq.~(\ref{eq:L-loop_ladder_diagram_def})
it will be useful to perform a change of variables which leaves
each soft propagator dependent on a single variable.
To this end, we adopt the change of variables introduced
in Ref.~\cite{Gardi:2011yz},
\begin{equation}
\left( \hspace{-1.1mm} \begin{array}{c} t_{\ell_j, m_j} \\
t_{r_j, n_j} \end{array} \hspace{-1.1mm} \right)
\hspace{0.2mm}=\hspace{0.2mm} \rho_j \left( \hspace{-1.1mm}
\begin{array}{c} x_j \\ 1 - x_j \end{array} \hspace{-1.1mm} \right)
\hspace{3mm} \mathrm{where} \hspace{3mm}
\left\{ \hspace{-0.6mm} \begin{array}{l}
0 \leq \rho_j < \infty \\[0.8mm]
0 \leq x_j \leq 1 \,.
\end{array} \right.
\label{eq:radial_coordinates}
\end{equation}
For notational convenience, we define the nesting function
in terms of the new variables as follows,
\begin{equation}
\Theta (\boldsymbol{\rho},\boldsymbol{x})
\hspace{-0.3mm}\equiv\hspace{-0.3mm} \prod_{i=1}^n \prod_{k=0}^{N_i}
\theta (t_{i,k+1} - t_{i,k}) \bigg|_{\tiny \left( \hspace{-1.5mm} \begin{array}{c} t_{\ell_j, m_j} \\
t_{r_j, n_j} \end{array} \hspace{-1.9mm} \right)
\hspace{0.7mm}=\hspace{0.7mm} \rho_j \left( \hspace{-1.3mm}
\begin{array}{c} x_j \\ 1 - x_j \end{array} \hspace{-1.7mm} \right)} \,,
\label{eq:nesting_function_def}
\end{equation}
and the soft propagators through
\begin{equation}
P_{ij}^{[\epsilon]} (x) \equiv \frac{v_i \cdot v_j}
{\big[ {-}\big(x v_i - (1-x)v_j\big)^2 + i\eta \big]^{1-\epsilon}} \,.
\label{eq:propagator_notation}
\end{equation}
The diagram then takes the form
\begin{equation}
F^{(L)} \hspace{-0.8mm}=\hspace{-0.3mm} \mu^{2L\epsilon} \prod_{j=1}^L
\int_0^\infty \hspace{-1.0mm} \frac{\d \rho_j}{\rho_j^{1-2\epsilon}} \hspace{-0.5mm}
\int_0^1 \hspace{-0.4mm} \d x_j \hspace{0.4mm} P_{\ell_j r_j}^{[\epsilon]} (x_j)
\Theta (\boldsymbol{\rho},\boldsymbol{x}) \,.
\label{eq:L-loop_ladder_diagram_rep_1}
\end{equation}
We observe that the dependence of the soft propagators on the
radial coordinates $\rho_j$ has scaled out, and that
each propagator now depends only on a single variable $x_j$.

Next, we extract the overall infrared divergence of the
diagram by setting $\tau_1 \equiv \rho_1$ and applying
the following sequence of $L-1$ substitutions
\begin{equation}
\left( \hspace{-1.1mm} \begin{array}{c} \tau_j \\
\rho_{j+1} \end{array} \hspace{-1.1mm} \right) =
\tau_{j+1} \left( \hspace{-1.1mm} \begin{array}{c} y_j \\
1 - y_j \end{array} \hspace{-1.1mm} \right) \hspace{4mm} \mathrm{with} \hspace{3mm}
\left\{ \hspace{-0.7mm} \begin{array}{l}
0\leq \tau_j < \infty \\[0.1mm]
0\leq y_j \leq 1 \,,
\end{array} \right.
\label{eq:tau_y_variables}
\end{equation}
where $j=1,\ldots,L-1$, the variables $\tau_j$
have the dimension of length, and the $y_j$ are dimensionless.

The $L$-loop eikonal diagram then becomes
\begin{equation}
F^{(L)} = \prod_{j=1}^L \int_0^1 \d x_j \hspace{0.5mm}
P_{\ell_j r_j}^{[\epsilon]} (x_j) \hspace{0.4mm} \mathcal{I} (\boldsymbol{x}) \,.
\label{eq:L-loop_ladder_diagram_rep_2}
\end{equation}
The infrared divergence of the diagram has been absorbed into the kernel
\begin{equation}
\mathcal{I} (\boldsymbol{x}) \hspace{-0.6mm}=\hspace{-0.2mm} \Gamma(2L\epsilon) \hspace{-0.5mm}
\left( \frac{\mu}{\Lambda} \right)^{2L\epsilon}
\prod_{j=1}^{L-1} \hspace{-0.4mm} \int_0^1 \hspace{-0.1mm}
\frac{\d y_j \hspace{0.8mm} y_j^{-1+2j\epsilon}}
{(1 \hspace{-0.2mm}-\hspace{-0.2mm} y_j)^{1-2\epsilon}}
\Theta \big(\{\boldsymbol{y}, \boldsymbol{x} \}\big),
\label{eq:kernel}
\end{equation}
where the notation $\Theta \big(\{\boldsymbol{y}, \boldsymbol{x} \}\big)$
refers to the result of applying the substitutions
(\ref{eq:tau_y_variables}) to Eq.~(\ref{eq:nesting_function_def}).
Here we have regulated the infrared divergence through
the damping factor $e^{-\Lambda \tau_L}$
with $\Lambda \ll 1$. In addition, Eq.~(\ref{eq:kernel}) contains any potential
ultraviolet subdivergences of the diagram.

Having written the $L$-loop eikonal diagram in the form
(\ref{eq:L-loop_ladder_diagram_rep_2}), we now turn to
the question of extracting its imaginary part. We will
restrict our attention to the leading order in the dimensional
regulator $\eps$ and drop the dependence of
the soft propagators on $\eps$,
\begin{equation}
F^{(L)} \hspace{-0.3mm}=\hspace{-0.3mm}
\prod_{j=1}^L \int_0^1 \d x_j \hspace{0.5mm}
P_{\ell_j r_j}^{[0]} (x_j) \hspace{0.4mm} \mathcal{I} (\boldsymbol{x})
\hspace{0.4mm} + \hspace{0.4mm} \mathcal{O}(\eps^{-d+1}) \,,
\label{eq:L-loop_ladder_diagram_rep_3}
\end{equation}
with $d$ denoting the degree of divergence of the diagram,
$F^{(L)} \sim \frac{1}{\eps^d} \times \mathrm{(finite)}$.

To compute the imaginary part of Eq.~(\ref{eq:L-loop_ladder_diagram_rep_3})
we start by observing that Eq.~(\ref{eq:kernel}) is
manifestly real. The Feynman $i\eta$'s are thus
the only source of imaginary parts of
Eq.~(\ref{eq:L-loop_ladder_diagram_rep_3}). We can therefore decompose each
of the $x_j$-integration paths into a principal-value part and
small semicircles around the propagator poles.
As the integrand takes purely imaginary values in the regions
close to the poles and is real-valued on the remaining domain of
integration, the resulting $2^L$ terms (each involving
$L$ integrations) will be either purely real or purely imaginary.

In order to collect the imaginary contributions, we define
the cut propagator
\begin{align}
\label{eq:Dij}
\Delta_{ij}(x) \equiv
- \pi \, v_i \cdot v_j \, \delta\big( (x v_i-(1-x) v_j)^2 \big) \,,
\end{align}
and the $p$-fold cutting operator
\begin{align}
\label{eq:Cut_x F}
\Cut_{x_{i_1},\dotsc,x_{i_p}} F^{(L)} &=
\prod_{j=1 \atop \!\!\! j \neq i_1,\dotsc,i_p \!\!\!\!\!\!}^{n} \PV \int_0^1 \d x_j \, P(x_j)\nonumber\\
&\hspace{6mm} \times \prod_{k=1}^{p} \int_0^1 \d x_{i_k} \,
\Delta(x_{i_k}) \hspace{0.3mm} \mathcal{I}(\boldsymbol{x})\,.
\end{align}
For notational brevity we omitted the indices on the (cut)
propagators: $P(x_j) \equiv P_{\ell_j r_j}^{[0]} (x_j)$
and $\Delta (x_j) \equiv \Delta_{\ell_j r_j} (x_j)$.

The imaginary part of any $L$-loop eikonal diagram with no
internal vertices is then, to the leading order in $\eps$,
\begin{equation}
\Im F^{(L)}=
\sum_{p=1 \atop p \hspace{0.7mm} \text{odd}}^L \hspace{0.0mm}
\sum_{i_1,\dotsc,i_p = 1 \atop i_1 < \dotsm < i_p}^L i^{\,p-1} \Cut_{x_{i_1}, \ldots, x_{i_p}} F^{(L)} \,.
\label{eq:Master Im F}
\end{equation}
The formula~(\ref{eq:Master Im F})
is illustrated for a generic ladder diagram in
Fig.~\ref{fig:Im_of_three-loop_ladder}.
Note that Eq.~(\ref{eq:Master Im F})
shows that the imaginary part of
the integrated result for the diagram
will have transcendentality weight one less than
the real part.

\begin{figure*}[!t]
\begin{center}
\includegraphics[width=0.90\textwidth]{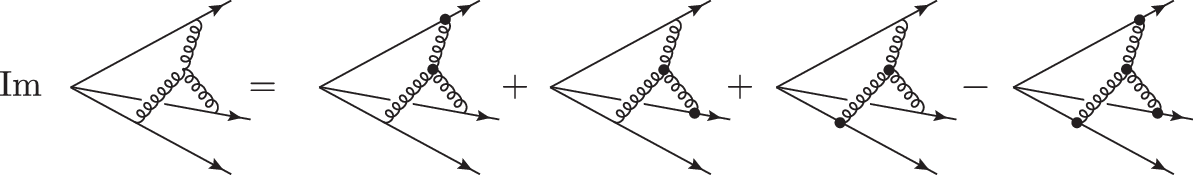}
{\vskip 1mm}
\caption{Schematic illustration of the formula~(\ref{eq:Master Im F})
for the imaginary part of an eikonal diagram with an internal vertex.}
\label{fig:Im_of_3g_vertex_diagram}
\end{center}
\end{figure*}

We have verified the formula~(\ref{eq:Master Im F})
for ladder-type diagrams with up to three loops,
finding agreement with results in the literature%
~\footnote{The diagram $F^{(3)}$ can be obtained from Ref.~\cite{Falcioni:2014pka}
as an appropriate linear combination of eqs.~(4.29) and (4.33) there.
Subsequent analytic continuation from spacelike to timelike kinematics
yields an imaginary part in agreement with Eq.~(\ref{eq:ImF3-v2}).}.
For example, for the diagram in Fig.~\ref{fig:Im_of_three-loop_ladder},
Eq.~(\ref{eq:Master Im F}) yields the imaginary part
($\gamma\equiv\gamma_{12}$, $\chi \equiv e^{-\gamma}$)
\begin{align}
\label{eq:ImF3-v2}
\Im F^{(3)} \hspace{-0.6mm}&=\hspace{-0.6mm}
\frac{\pi}{6\eps} \left( \frac{\mu}{\Lambda}
\right)^{6\eps} \hspace{-1.0mm} \coth^3 \hspace{-0.9mm} \gamma \hspace{0.3mm}
\Big[H_{3,1}(\chi^2) \hspace{-0.2mm}+\hspace{-0.2mm}H_{2,2}(\chi^2)
\hspace{-0.2mm}+\hspace{-0.2mm}H_4(\chi^2) \nonumber \\
&\hspace{-8mm} {-}\tfrac{1}{3}\log^4 \chi +\log^2 \hspace{-0.4mm} \chi
\hspace{0.4mm} \big(H_2(\chi^2) \hspace{-0.2mm}+\hspace{-0.2mm} 3\zeta_2\big)
-\log \chi \big(H_3(\chi^2) \hspace{-0.2mm} - \hspace{-0.2mm} \zeta_3\big) \nonumber\\
&\hspace{-8mm}
-\zeta_2 H_2(\chi^2)
+\tfrac{1}{2}\zeta_4
\Big] + \mathcal{O}(\eps^0)
\,.
\end{align}
The $H_{i,j}$ and $H_i$ denote harmonic polylogarithms
according to the conventions of Ref.~\cite{Remiddi:1999ew}.

A natural question concerns the relation of the imaginary part
of the eikonal diagram to the discontinuities in its various
kinematic channels. Expressed in terms of the \emph{exponentials}
of the cusp angles, $\chi_{ij} \equiv e^{-\gamma_{ij}}$,
rather than the cusp angles $\cosh \gamma_{ij} = |v_i \cdot v_j|$
themselves, the eikonal diagram has branch cuts located on the real line and
satisfies Schwarz reflection, $F^{(L)}(\overline{\chi_{ij}})
= \overline{F^{(L)}(\chi_{ij})}$. As a result, the discontinuities
give rise to the imaginary part through the relation
\begin{equation}
2i \hspace{0.7mm} \mathrm{Im} \hspace{0.6mm}
F^{(L)} (\boldsymbol{\chi})
\hspace{0.7mm}=\hspace{0.7mm} \sum_{j=1}^L \theta(v_{\ell_j} \cdot v_{r_j}) \hspace{0.2mm}
\mathop{\mathrm{Disc}}_{\chi_{\ell_j r_j}} \hspace{0.3mm}
F^{(L)} (\boldsymbol{\chi}) \,.
\label{eq:Im_and_discontinuities_relation}
\end{equation}
The step functions account for the fact that the
imaginary part has non-vanishing and vanishing contributions
from channels with, respectively, timelike ($v_{\ell_j} \cdot v_{r_j} > 0$)
and spacelike ($v_{\ell_j} \cdot v_{r_j} < 0$) kinematics,
in agreement with the causality considerations of the previous
section.

\section{Cuts of eikonal diagrams with internal vertices}

Eikonal diagrams with internal vertices have a more
complicated structure than ladder-type diagrams. Nonetheless,
we expect that our cutting prescription applies to such
diagrams as well. As an explicit example, we consider
the two-loop three-line diagram involving
a three-gluon vertex, illustrated in Fig.~\ref{fig:Im_of_3g_vertex_diagram}.
In order to apply our prescription,
the overall infrared divergence of the diagram is extracted
by integrating out the radial distance of the three-gluon
vertex, leaving three remaining integrations over real
projective space $u \in \mathds{R}\mathds{P}^{1,2}$.
After some algebraic manipulations, the three-gluon vertex
diagram has the following position-space
representation, to the leading order in $\epsilon$ \cite{Mitov:2009sv},
\begin{align}
F_{3g}
\propto \frac{1}{\eps}
\int \d^3 u
~{\cal V}(\{v_i\},u)
\hspace{0.4mm} \prod_{\ell=1}^{3} \int_{0}^{\infty} \d x_\ell \,
P_{\ell}^{[0]}(\zeta_\ell, x_\ell) \,,
\label{eq:Three_gluon_vertex_diagram}
\end{align}
with the three-gluon vertex related differential operator
${\cal V}(\{v_i\},u)=\sum_{i,j,k=1}^{3} \varepsilon_{ijk}
\, v_i \cdot v_j \, \zeta_i \, \zeta_k \, \frac{\partial}{\partial \zeta_i}$,
in terms of the scalar products $\zeta_i = v_i \cdot u$.
The position-space propagators are given by
$P_{i}^{[\eps]}(\zeta_i,x_i) = (-u^2 + 2 x_i \zeta_i - x_i^2 + i\eta)^{-1+\eps}$.
The imaginary part of this diagram is given by the formula~(\ref{eq:Master Im F})
with the cut propagators,
\begin{align}
\Delta_{i}(x) \equiv
- \pi \, \delta\big(u^2 - 2 x_i \zeta_i + x_i^2 \big) \,.
\end{align}
The formula is illustrated in Fig.~\ref{fig:Im_of_3g_vertex_diagram}.
We have checked the formula by performing the integral over the
direction of the three-gluon vertex $u$ numerically.
For small cusp angles, where the numerics proves to behave well,
we find agreement with the analytic result in Ref.~\cite{Ferroglia:2009ii}.
This in turn suggests the applicability of our cutting
prescription to any eikonal diagram.

{\bf Acknowledgments}

We thank S.~Abreu, S.~Caron-Huot, E.~Gardi, J.~Henn, P.~Hoyer,
L.~Magnea, G.~Sterman, I.~Stewart, W.~Vleeshouwers, A.~Waelkens, C.~White
and especially G.~Korchemsky for useful discussions.
KJL is grateful for the hospitality of
the Institut de Physique Th{\'e}orique, CEA Saclay,
where part of this work was carried out.
The research leading to these results has received
funding from the European Union Seventh Framework
Programme (FP7/2007-2013) under grant agreement no.~627521.
This work was supported by the Foundation for
Fundamental Research of Matter (FOM),
program 104 ``Theoretical Particle Physics in the Era of the LHC''
and by the Research Executive Agency (REA)
of the European Union under the Grant Agreement number
PITN-GA-2010-264564 (LHCPhenoNet).

\bibliography{Wilson_line_cuts}

\end{document}